\documentclass{article}
\usepackage{spconf,amsmath,graphicx,hyperref}
\usepackage{amsmath}
\usepackage{booktabs}
\usepackage{xcolor}
\usepackage{amssymb}
\usepackage{subcaption}

\title{Listening for Airway Stenosis: A Foundation Model-Based Method for Rapid and Accessible Detection}
\name{Jean Groeninger\thanks{$\star$ Equal contribution.}$^{1,2,\star}$, Zihao Zhao$^{1,\star}$, Juliana de Castilhos$^{1}$, Sven Nebelung$^{1}$, Daniel Truhn$^{1}$}
\address{
$^{1}$ Department of Diagnostic and Interventional Radiology, University Hospital Aachen\\$^{2}$ Télécom Paris }

\begin{document}
%
\maketitle
\begin{abstract}
Airway stenosis can cause severe respiratory complications, yet its detection often relies on specialized examinations and medical imaging. This study explores the potential of acoustic AI for rapid and accessible airway stenosis detection using readily acquired patient voice recordings. We systematically investigate whether acoustic foundation models (AFMs) can extract acoustic representations associated with airway stenosis-related speech patterns.
Experiments are conducted on a cohort of 748 participants from the Bridge2AI-Voice dataset, 134 with airway stenosis and 614 without. 
The best-performing model achieves an AUROC of 0.952 and an accuracy of 0.924 (means over five-fold cross-validation), highlighting the potential of AFMs to transfer beyond general-purpose speech applications to clinical diagnostic tasks. Further analysis reveals that the model primarily relies on connected-speech recordings rather than isolated acoustic tasks, such as sustained phonation and breathing. Overall, these results suggest that voice-based acoustic AI could complement existing diagnostic workflows by enabling rapid, low-burden, and widely accessible screening for airway stenosis. 
Code is available at
\href{https://github.com/JeanGrng/listening-for-airway-stenosis}{Github}.
\end{abstract}
\section{Introduction}\label{sec:intro}

Airway stenosis is an abnormal narrowing of the laryngeal or tracheal airway and may lead to progressive dyspnea, stridor, and potentially life-threatening respiratory compromise. Early airway stenosis often presents with nonspecific symptoms and can be mistaken for more common respiratory disorders, which may delay diagnosis and treatment~\cite{ntouniadakis2022monitoring}. Clinical evaluation typically combines endoscopic examinations, such as laryngoscopy or bronchoscopy, with computed tomography and pulmonary function testing. These tests are important for diagnosis, but they require medical equipment, trained staff, and more time and resources than voice recording. They are therefore less practical for frequent, rapid, or remote assessment.

\begin{figure*}
    \centering
    \includegraphics[width=0.95\textwidth]{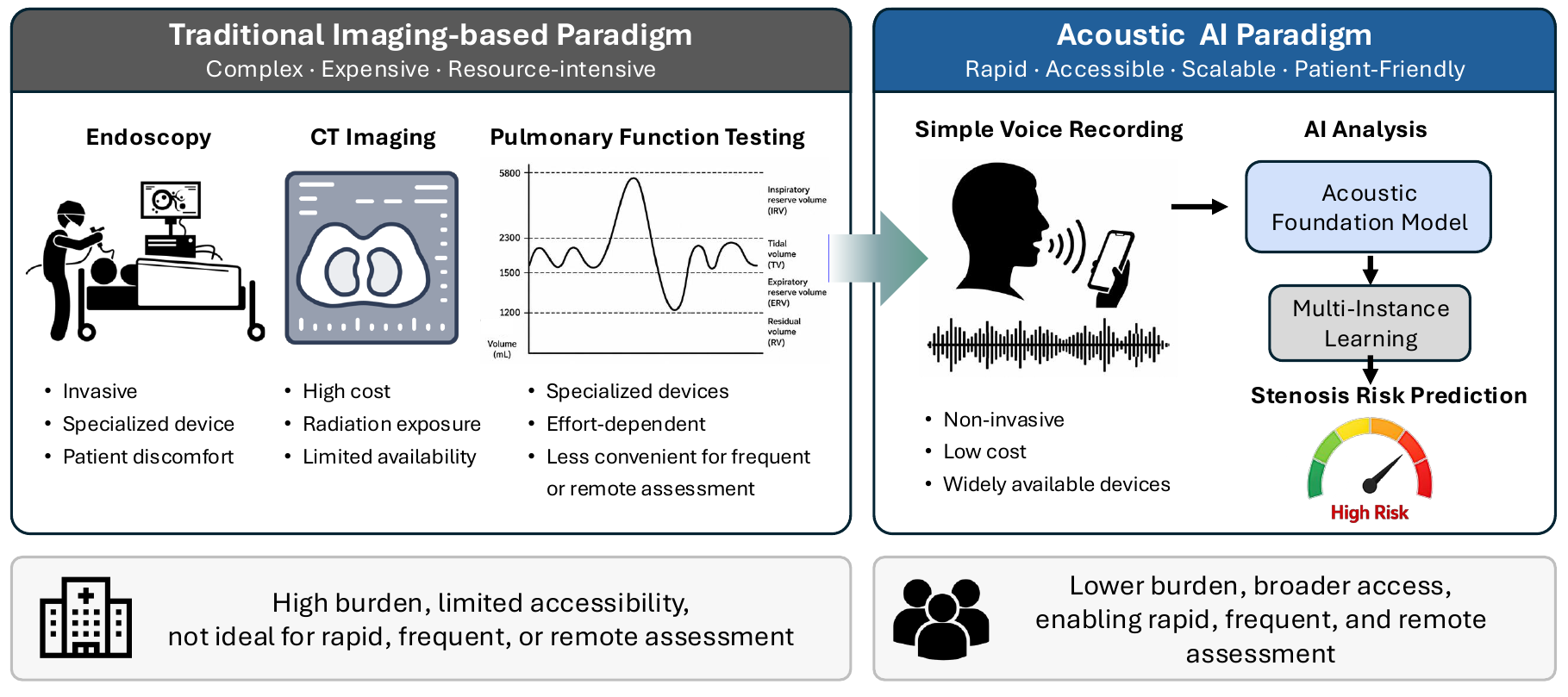}
    \caption{Overview of the paradigm shift from traditional airway stenosis assessment to acoustic AI-based detection.
While conventional approaches rely on specialized examinations and resource-intensive workflows, voice-based acoustic AI enables rapid and accessible assessment by analyzing patient voice recordings with acoustic foundation models and patient-level aggregation.
}
    \label{fig:teaser}
\end{figure*}
Voice is produced by airflow from the lungs, vibration of the vocal folds, and filtering by the vocal tract. When the upper airway becomes narrow, airflow and pressure during speech can change. These changes may affect voice quality, noise, and resonance. Previous studies have reported abnormal voice characteristics in patients with subglottic stenosis and have shown that severe stenosis can alter airflow and voice production~\cite{hilton2025aerodynamic}. However, these changes are not always the same across patients and may be difficult to describe with a small number of conventional acoustic features.
Modern acoustic models can learn directly from voice recordings and may capture patterns that are missed by traditional handcrafted features. Voice recording is also fast, simple, and can be implemented via standard consumer-grade devices. This makes acoustic AI a potential complementary tool with improved accessibility for airway stenosis detection.

Acoustic foundation models (AFMs) are pretrained on large-scale speech or general-audio corpora. They have demonstrated strong transferability across speech recognition, paralinguistic analysis, and audio classification tasks~\cite{hsu2021hubert,chen2022wavlm,chen2023beats}. Their pretrained representations are therefore attractive for clinical applications, where large-scale labeled datasets are usually difficult to collect.  
However, AFMs differ significantly in their pretraining data, objectives, and architectures.
Also, previous clinical speech analysis studies mostly applied AFMs to neurological diseases and motor speech disorders~\cite{chen2023exploring,wiepert2024speech}, while airway stenosis is a structural airway disease whose acoustic manifestations may arise from airflow restriction, turbulence, and altered resonance. Although Anibal et al.~\cite{anibal2025transformers} did a pilot study by using YAMNet~\cite{gemmeke2017audio} as a frozen acoustic feature extractor, systematic comparisons across different AFMs and strategies for leveraging their representations remain largely unexplored.

In this work, we investigate the potential of voice-based acoustic AI for rapid and accessible airway stenosis detection using the Bridge2AI-Voice cohort~\cite{PhysioNet-b2ai-voice-3.0.0}. We systematically evaluate multiple AFMs by training diagnostic classifiers on their frozen acoustic representations. 
Our study addresses three questions: (1) whether patient voice recordings contain sufficient information to distinguish individuals with airway stenosis from controls, (2) how the choice of AFM influences performance, and (3) which speech tasks contribute most to patient-level prediction. 
Our results suggest that voice-based acoustic analysis may complement current diagnostic methods as a fast and simple assessment tool.

\section{Method}\label{sec:method}
In this section, we first formulate the problem we addressed in this study. We then describe how AFMs are used to extract task-level representations from each recording. Finally, we introduce the multiple-instance learning framework that aggregates these representations into a patient-level prediction.

\subsection{Problem formulation}
Our framework formulates airway stenosis detection as a patient-level classification problem from multiple acoustic recordings. Each patient completes a set of predefined voice and speech tasks, and the resulting recordings are treated jointly rather than as independent samples. 
Let the dataset contain $N$ patients,
\[
\mathcal{D} = \{(\mathcal{X}_i, y_i)\}_{i=1}^{N},
\]
where $y_i \in \{0,1\}$ denotes the absence or presence of airway stenosis. Unlike conventional audio classification, each patient is represented by multiple acoustic recordings,
\[
\mathcal{X}_i =
\{x_i^{(1)}, x_i^{(2)}, \ldots, x_i^{(M_i)}\},
\]
corresponding to different voice or speech tasks. The number of available recordings $M_i$ may vary across patients.
The objective is therefore not to classify an individual recording, but to infer a single patient-level detection from the complete set of available recordings:
\[
\hat{y}_i = F(\mathcal{X}_i).
\]

\subsection{Acoustic foundation model representations}

Let $f_{\theta}$ denote the AFM, each recording $x_i^{(m)}$ is passed independently through the frozen encoder to obtain a sequence of latent acoustic features,
\[
\mathbf{H}_i^{(m)}
=
f_{\theta}\!\left(x_i^{(m)}\right)
\in
\mathbb{R}^{T_m \times d},
\]
where $T_m$ denotes the number of latent frames and $d$ is the representation dimension. Temporal mean pooling is then applied to obtain one fixed-dimensional embedding:
\[
\mathbf{z}_i^{(m)}
=
\frac{1}{T_m}
\sum_{t=1}^{T_m}
\mathbf{H}_{i,t}^{(m)}.
\]
The patient is therefore represented as a bag of task-level embeddings,
\[
\mathcal{B}_i =
\left\{
\mathbf{z}_i^{(1)},
\mathbf{z}_i^{(2)},
\ldots,
\mathbf{z}_i^{(M_i)}
\right\}.
\]
We evaluate multiple pretrained AFMs under the same protocol. This allows us to examine whether representations learned from large-scale general-purpose speech and audio corpora transfer differently to airway stenosis detection, while avoiding confounding effects from task-specific fine-tuning.

\subsection{Patient-level multiple-instance aggregation}
The task embeddings within $\mathcal{B}_i$ are assumed to contain complementary information from different speaking conditions. We therefore aggregate them before making a patient-level prediction.
We use TransMIL~\cite{shao2021transmil} as our primary aggregation module, which models interactions across task-level embeddings. Given the patient bag $\mathcal{B}_i$, the MIL module produces a patient representation
\[
\mathbf{h}_i = g_{\phi}(\mathcal{B}_i),
\]
which is passed to a binary classification head:
\[
p_i =
\sigma\!\left(
\mathbf{w}^{\top}\mathbf{h}_i + b
\right),
\]
where $p_i$ denotes the predicted probability of airway stenosis. In this study, we select WavLM Layer 15 representations combined with TransMIL aggregation as the default setting.

\begin{table}[t]
\centering
\caption{Comparison of different feature extraction methods for airway stenosis detection. 
For each model, we report the performance over five folds.}
\label{tab:fm_comparison}
\small
\setlength{\tabcolsep}{5pt}
\begin{tabular}{lccc}
\toprule
\textbf{Method} & \textbf{AUROC} & \textbf{F1-score} & \textbf{Accuracy} \\
\midrule

Mel + MFCC + F0
& $0.792_{\scriptscriptstyle \pm 0.030}$
& $0.484_{\scriptscriptstyle \pm 0.073}$
& $0.767_{\scriptscriptstyle \pm 0.049}$ \\

\midrule
\multicolumn{4}{l}{\textit{General audio}} \\

AST~\cite{gong2021ast}
& $0.886_{\scriptscriptstyle \pm 0.017}$
& $0.579_{\scriptscriptstyle \pm 0.045}$
& $0.861_{\scriptscriptstyle \pm 0.014}$ \\

AudioMAE~\cite{huang2022masked}
& $0.806_{\scriptscriptstyle \pm 0.042}$
& $0.479_{\scriptscriptstyle \pm 0.083}$
& $0.799_{\scriptscriptstyle \pm 0.033}$ \\

OpenBEATs~\cite{bharadwaj2025openbeats}
& $0.751_{\scriptscriptstyle \pm 0.054}$
& $0.444_{\scriptscriptstyle \pm 0.082}$
& $0.780_{\scriptscriptstyle \pm 0.052}$ \\

SSAST~\cite{gong2022ssast}
& $0.818_{\scriptscriptstyle \pm 0.032}$
& $0.498_{\scriptscriptstyle \pm 0.046}$
& $0.832_{\scriptscriptstyle \pm 0.011}$ \\

YAMNet~\cite{gemmeke2017audio}
& $0.816_{\scriptscriptstyle \pm 0.047}$
& $0.529_{\scriptscriptstyle \pm 0.061}$
& $0.822_{\scriptscriptstyle \pm 0.030}$ \\

\midrule
\multicolumn{4}{l}{\textit{Speech-specialized}} \\

HuBERT~\cite{hsu2021hubert}
& $0.943_{\scriptscriptstyle \pm 0.032}$
& $0.728_{\scriptscriptstyle \pm 0.062}$
& $0.894_{\scriptscriptstyle \pm 0.026}$ \\

Whisper~\cite{radford2023robust}
& $0.925_{\scriptscriptstyle \pm 0.009}$
& $0.692_{\scriptscriptstyle \pm 0.037}$
& $0.888_{\scriptscriptstyle \pm 0.024}$ \\

\textbf{WavLM}~\cite{chen2022wavlm}
& $\mathbf{0.952}_{\scriptscriptstyle \pm 0.033}$
& $\mathbf{0.783}_{\scriptscriptstyle \pm 0.077}$
& $\mathbf{0.924}_{\scriptscriptstyle \pm 0.026}$ \\

\bottomrule
\end{tabular}
\end{table}
\section{Results}
\label{sec:res}

\subsection{Experimental setting}

We evaluate our framework on the Bridge2AI-Voice dataset~\cite{PhysioNet-b2ai-voice-3.0.0,pollard2026physionet}. Our cohort comprises 748 participants, 134 diagnosed with airway stenosis and 614 without, each contributing multiple recordings collected from 16 different task groups at a sampling rate of 16 kHz. The negative class is not restricted to healthy speakers: it comprises all participants without an airway-stenosis label, most of whom present other laryngeal, neurological, psychiatric or respiratory conditions. 
The task groups can be further categorized into connected speech, simple diadochokinetic (DDK), sustained acoustic, and
respiration. 
Following the patient-level formulation, all recordings from the same participant are jointly used for detection. 
The dataset provides linear magnitude spectrograms rather than waveforms. Models that operate on raw audio therefore receive a Griffin-Lim reconstruction.
We perform five-fold cross-validation with patient-level splitting. 
Model selection and performance estimation used the same cross-validation framework. Reported results therefore represent cross-validated development performance.
All experiments are conducted using a single NVIDIA L40S GPU.

\begin{table}[t]
\centering
\caption{Ablation studies of the proposed framework.
(a) Effect of WavLM layer selection on  representation quality.
(b) Comparison of different patient-level aggregation strategies.}
\label{tab:ablation}
\small

\begin{subtable}{\linewidth}
\centering
\caption{Layer selection.}
\label{tab:wavlm_layer_ablation}
\setlength{\tabcolsep}{6pt}
\begin{tabular}{lccc}
\toprule
\textbf{Layer} & \textbf{AUROC} & \textbf{F1-score} & \textbf{Accuracy} \\
\midrule
L6
& $0.930_{\scriptscriptstyle \pm 0.030}$
& $0.697_{\scriptscriptstyle \pm 0.057}$
& $0.884_{\scriptscriptstyle \pm 0.033}$ \\

L10
& $0.942_{\scriptscriptstyle \pm 0.033}$
& $0.705_{\scriptscriptstyle \pm 0.052}$
& $0.893_{\scriptscriptstyle \pm 0.017}$ \\

\textbf{L15}
& $\mathbf{0.952}_{\scriptscriptstyle \pm 0.033}$
& $\mathbf{0.783}_{\scriptscriptstyle \pm 0.077}$
& $\mathbf{0.924}_{\scriptscriptstyle \pm 0.026}$ \\

L20
& $0.926_{\scriptscriptstyle \pm 0.038}$
& $0.720_{\scriptscriptstyle \pm 0.063}$
& $0.886_{\scriptscriptstyle \pm 0.033}$ \\

L24
& $0.923_{\scriptscriptstyle \pm 0.038}$
& $0.728_{\scriptscriptstyle \pm 0.068}$
& $0.896_{\scriptscriptstyle \pm 0.038}$ \\

\bottomrule
\end{tabular}
\end{subtable}

\vspace{0.5em}

\begin{subtable}{\linewidth}
\centering
\caption{Aggregation strategy.}
\label{tab:aggregation_ablation}
\setlength{\tabcolsep}{6pt}
\begin{tabular}{lccc}
\toprule
\textbf{Aggregation} & \textbf{AUROC} & \textbf{F1-score} & \textbf{Accuracy} \\
\midrule

Non-MIL
& $0.932_{\scriptscriptstyle \pm 0.041}$
& $0.712_{\scriptscriptstyle \pm 0.059}$
& $0.893_{\scriptscriptstyle \pm 0.033}$ \\



GatedAttn~\cite{ilse2018attention}
& $0.945_{\scriptscriptstyle \pm 0.041}$
& $0.762_{\scriptscriptstyle \pm 0.045}$
& $0.906_{\scriptscriptstyle \pm 0.026}$ \\

DSMIL~\cite{li2021dual}
& $0.944_{\scriptscriptstyle \pm 0.038}$
& $0.723_{\scriptscriptstyle \pm 0.049}$
& $0.892_{\scriptscriptstyle \pm 0.029}$ \\

\textbf{TransMIL}~\cite{shao2021transmil}
& $\mathbf{0.952}_{\scriptscriptstyle \pm 0.033}$
& $\mathbf{0.783}_{\scriptscriptstyle \pm 0.077}$
& $\mathbf{0.924}_{\scriptscriptstyle \pm 0.026}$ \\

\bottomrule
\end{tabular}
\end{subtable}

\end{table}
\subsection{Comparison between different AFMs}
Table~\ref{tab:fm_comparison} summarizes the performance of different pretrained acoustic representations for airway stenosis detection. Among all evaluated models, speech-specialized foundation models achieve the strongest performance, with WavLM reaching an AUROC of $0.952_{\scriptscriptstyle \pm 0.033}$.
The three speech-specialized models, WavLM, HuBERT, and Whisper, all achieve AUROC values above 0.92. In comparison, general-audio models perform worse, with the best model reaching an AUROC of $0.886_{\scriptscriptstyle \pm 0.017}$. A conventional handcrafted baseline, combining log-mel, MFCC and F0 statistics, achieves an AUROC of only $0.792_{\scriptscriptstyle \pm 0.030}$.
These results suggest that, under our controlled setting, pretrained speech models transfer better to airway stenosis detection than general-audio models or handcrafted acoustic features.

\begin{figure}
    \centering
    \includegraphics[width=0.5\textwidth]{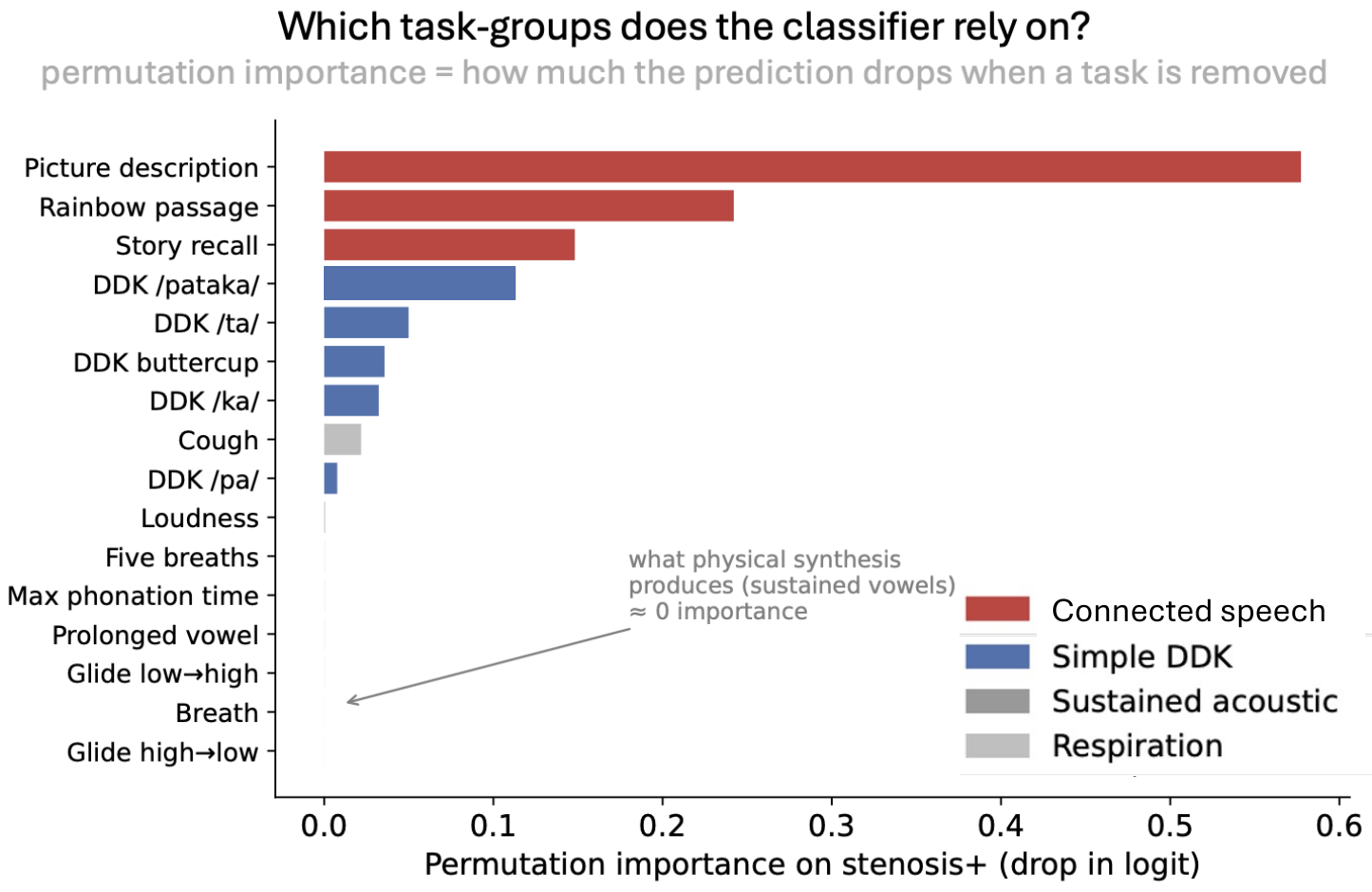}
    \caption{
Task-level contribution analysis for airway stenosis detection.
Connected-speech tasks provide the largest contributions, whereas sustained phonation and breathing-related tasks contribute minimally.
}
    \label{fig:task_importance}
\end{figure}
\subsection{Ablation study}
\label{sec:ablation}
We further investigate the influence of representation depth and patient-level aggregation strategy using WavLM. As shown in Table~\ref{tab:ablation}(a), the diagnostic performance varies across transformer layers. Intermediate layers provide the most discriminative representations, with Layer 15 achieving the highest AUROC of $0.952_{\scriptscriptstyle \pm 0.033}$. Earlier and later layers show slightly reduced performance, suggesting that mid-level layers may preserve more related information.

Table~\ref{tab:ablation}(b) compares different patient-level aggregation strategies using WavLM Layer 15 representations. A single recording already provides strong diagnostic information, achieving an AUROC of $0.932_{\scriptscriptstyle \pm 0.041}$. However, modeling multiple recordings jointly further improves performance, especially in F1-score and Accuracy. Among the evaluated aggregation strategies, TransMIL achieves the best performance, outperforming non-MIL and alternative MIL approaches, with an AUROC of $0.952_{\scriptscriptstyle \pm 0.033}$. This indicates that explicitly modeling the complementary information across heterogeneous speech tasks benefits patient-level detection.

\subsection{Analysis of task-level contributions}

To investigate which acoustic tasks contribute most to the final prediction, we perform a leave-one-task-out permutation analysis on the best-performing model (WavLM Layer 15 with TransMIL aggregation). For each task group, the corresponding task embedding is replaced by the mean embedding estimated from stenosis-negative training participants, and the resulting change in the pre-sigmoid prediction logit is used as the importance measure. This in-distribution perturbation strategy avoids potential artifacts caused by replacing representations with unrealistic values~\cite{sturmfels2020visualizing}.

As shown in Fig.~\ref{fig:task_importance}, the predictive information is primarily concentrated in connected-speech tasks. Picture description contributes the largest prediction change, followed by Rainbow Passage, story recall, and the DDK /pataka/ task. In contrast, sustained phonation tasks, including prolonged vowels and glides, as well as breathing-related tasks, show minimal contribution. These findings suggest that the model primarily exploits speech production patterns emerging during natural connected speech rather than isolated acoustic measurements.
Importantly, the task contribution is disease-specific. 
The same perturbations applied to stenosis-negative participants produce substantially smaller changes in model predictions. This contrast indicates that the identified importance patterns are more closely related to stenosis than to general sensitivity to particular recording conditions. Respiratory task groups emphasized in previous voice-based airway studies~\cite{anibal2025transformers} show relatively little contribution. Instead, the foundation model appears to rely on information distributed across other voice tasks, which may not be captured by explicitly designed respiratory measurements.

\begin{table}[t]
\centering
\caption{Performance on phenotyping tasks beyond binary airway stenosis detection, using WavLM L15 with TransMIL.}
\label{tab:phenotyping}
\small
\begin{tabular}{lc}
\toprule
\textbf{Task} & \textbf{macro-averaged AUROC} \\
\midrule
Localization (3-class) & $0.820_{\scriptscriptstyle \pm 0.060}$ \\
Stridor detection (binary) & $0.669_{\scriptscriptstyle \pm 0.189}$ \\
Severity grading (3-class) & $0.604_{\scriptscriptstyle \pm 0.063}$ \\
\bottomrule
\end{tabular}
\end{table}

\subsection{Beyond binary detection}
While the proposed framework performs well for binary airway stenosis detection, we further evaluated three clinically relevant phenotyping tasks: anatomical localization of stenosis, detection of stridor, and severity grading. 
Unlike binary detection, these tasks test whether the learned acoustic features can capture differences between patients with airway stenosis. As shown in Table ~\ref{tab:phenotyping}, localization achieves moderate performance, whereas stridor detection and severity grading are more difficult. This suggests that voice signals contain clearer information about whether stenosis is present than about its specific clinical features. More reliable phenotype prediction will likely require larger cohorts that include a wider range of disease presentations.

\section{Conclusion and Discussion}

In this work, we investigated the potential of voice-based acoustic AI for rapid and accessible airway stenosis detection. By leveraging pretrained AFMs and patient-level multiple-instance learning, our framework effectively identified patients with airway stenosis from readily acquired voice recordings. 
Further analysis indicated that the model primarily relied on connected-speech tasks rather than isolated phonation or breathing recordings, suggesting that clinically relevant airway information may be embedded in complex speech production patterns.
Exploratory phenotyping further revealed that finer-grained characteristics, particularly stenosis severity, remain challenging to predict. As this study was conducted on a single cohort without external validation, the reported performance should be interpreted as a development-stage estimate. Future studies should validate these findings in larger and more diverse clinical cohorts.

{
\noindent\textbf{Compliance with Ethical Standards:}
This research study was conducted retrospectively using de-identified human subject data made available by the Bridge2AI-Voice consortium through PhysioNet under credentialed access. Ethical approval was not required as confirmed by the license and data use agreement attached with the data.
}
\bibliographystyle{IEEEbib}
\bibliography{refs}

\end{document}